\documentclass[prl,superscriptaddress,showpacs,twocolumn]{revtex4}
\usepackage{graphicx}
\usepackage{graphicx,amsfonts}
\usepackage{epsfig,amsmath}
\usepackage{verbatim}
\def\figwidth{8cm}

\begin{document}

\newcommand{\atanh}
{\operatorname{atanh}}
\newcommand{\ArcTan}
{\operatorname{ArcTan}}
\newcommand{\ArcCoth}
{\operatorname{ArcCoth}}
\newcommand{\Erf}
{\operatorname{Erf}}
\newcommand{\Erfi}
{\operatorname{Erfi}}
\newcommand{\Ei}
{\operatorname{Ei}}

\title{Functional Renormalization for 
pinned elastic systems away from their steady states}
\author{Gregory Schehr}
\affiliation{Theoretische Physik Universit\"at des Saarlandes
66041 Saarbr\"ucken Germany}
\author{Pierre Le Doussal}
\affiliation{LPTENS CNRS UMR 8549 24, Rue Lhomond 75231 Paris
Cedex 05, France}
\date{\today}

\begin{abstract}
Using one loop functional RG we study two problems
of pinned elastic systems away from their
equilibrium or steady states.
The critical regime of the depinning transition is investigated
starting from a flat initial condition. It
exhibits non trivial two-time dynamical regimes
with exponents and scaling functions obtained
in a dimensional expansion. The aging and equilibrium dynamics 
of the super-rough glass phase of the random Sine-Gordon model 
at low temperature is found to be characterized by a single
dynamical exponent $z \approx c/T$, where $c$ compares
well with recent numerical work. This agrees 
with the thermal boundary layer picture
of pinned systems.
\end{abstract}
\pacs{}
\maketitle

Disordered elastic systems offer many experimental
realizations and are also of theoretical
interest as prototype models for glasses induced by
quenched disorder. The competition between the structural
order and substrate impurities results in pinning, 
complex ground states, barriers and
ultra slow glassy dynamics. While ground state, equilibrium dynamics, 
and driven steady state properties have been much 
studied theoretically, much less is known about the dynamics {\it before} the
steady state is reached, or about the aging dynamics.
If universality is shown there, it would be of high interest 
for numerous experimental systems, e.g. magnetic domain wall
relaxation \cite{creepexp}, superconductors \cite{vortices}, contact
line depinning \cite{rolley}, density waves~\cite{cdw}. 

Numerical studies of glassy dynamics are hampered
by high barriers in configuration space resulting in
ultra-long time scales making comparison with 
theory uncertain. In some cases however
faster, but still interesting dynamics occurs.
One is zero temperature $T=0$ driven dynamics near the
depinning transition where barriers disappear \cite{frgdep1}.
Recent theoretical progress has been achieved there.
Functional renormalization 
group (FRG) studies give more precise and consistent
predictions \cite{frgdep2}, corroborated by powerful new algorithms 
which allow for excellent determination of steady state exponents
\cite{rosso, allemands}. 
One aim of this paper is to extend the FRG to dynamics 
away from steady states, and to show that interesting 
universal two-time dynamics (analogous to
the aging dynamics in non driven situations)
also occurs near depinning. We predict new exponents
and scaling functions in the critical regime. 

A second case where $T>0$ dynamics has been
investigated is the ''marginal glass'' phase exhibited
by topologically ordered 2D periodic systems 
as captured by the Cardy Ostlund (CO) model \cite{co}.
Barriers there grow only logarithmically
with size allowing for precise numerics \cite{simu_tc}. The 
equilibrium \cite{tsai_super_rough} and aging
\cite{schehr_co_pre} dynamics were studied  
using Coulomb gas RG methods, but only near the glass transition
temperature $T_g$ \cite{toner}. A numerical study has
confirmed some of the RG predictions, 
and in addition has explored the full temperature
regime \cite{schehr_co_num}. The dynamical exponent was found
to diverge as $z \sim 1/T$ at low $T$. One aim of this paper
is to show this result within
a simple one loop FRG, and to obtain detailed predictions
for the aging regimes at low $T$. This study,
together with the rather good agreement with numerics,
is also important as indirect evidence for
a recent hypothesis that a ''thermal boundary layer'' (TBL)
in the field theory controls the activated dynamics
of (more strongly) pinned manifolds \cite{balentspld}. 
\begin{figure}[!h]
\hspace*{0.2cm}
\centerline{\includegraphics[width=9cm]{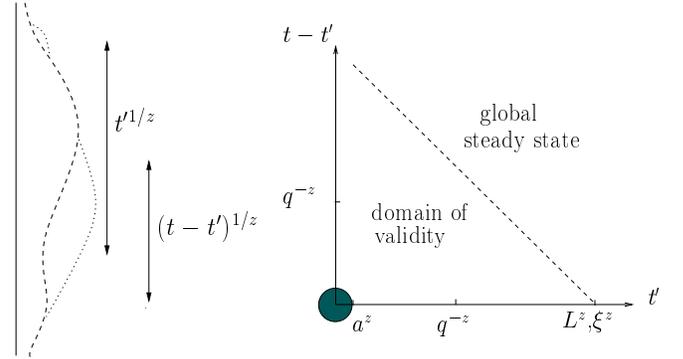}}

\caption{(a) solid line: initial flat
 configuration, dashed and dotted lines: configurations of
 the string at time $t'$ and $t$ respectively. (b)
Domain of validity, in logarithmic scale, of our approach near the
 depinning transition. 
The grey circle represents the small time region sensitive to
microscopic details.}
\end{figure}

The overdamped dynamics of a single component
elastic manifold of internal dimension $d$, 
parametrized by a height field $u(x,t)\equiv u_{xt}$
is described by
\begin{equation} \label{eq_dyn}
 \eta \partial_ t u_{x,t} = c \nabla^2 u_{x,t} + F(x, u_{x,t} + v t )
+ f - \eta v  + \xi(x,t) 
\end{equation}
where $\overline{u_{xt}}=0$,
$\langle \xi(x,t) \xi(x',t') \rangle = 2 T \eta \delta^d(x-x') \delta(t-t')$
is the correlator of the thermal gaussian noise, 
$\overline{F(x, u) F(x', u')}= \Delta(u-u') \delta^d(x-x')$
the second cumulant of the quenched random pinning force and 
$v$ is the average velocity. We denote the small scale cutoff size $a
\sim \Lambda_0^{-1}$. 
In this paper we consider a flat initial configuration $u_{xt=0}=0$.
Denoting $u_{qt}$ the spatial Fourier transform of $u_{x,t}$, 
we focus on the correlation ${\cal
C}^q_{tt'}$ and the connected (w.r.t. the thermal fluctuations) 
correlation ${\tilde{\cal C}}^q_{tt'}$: 
\begin{equation}\label{def_C}
{\cal{C}}^q_{tt'} = \overline{\langle {u}_{qt}
      {u}_{-qt'}  \rangle} \quad , \quad 
{\tilde{\cal C}}^q_{tt'} = \overline{\langle {u}_{qt}
      {u}_{-qt'}  \rangle} - 
\overline{{\langle {u}_{qt}\rangle} {\langle
      {u}_{-qt'}\rangle}}  
\end{equation}
and the response ${\cal R}^q_{tt'}$ to a small external
field ${h}_{-qt'}$  
\begin{equation}\label{def_R}
{\cal R}^q_{tt'} = \overline{ \delta \langle
  {u}_{qt} \rangle/\delta 
  {h}_{-qt'}}  \quad, \quad t > t'
\end{equation}
where $\overline{..}$ and $\langle .. \rangle $ denote respectively
averages w.r.t. disorder and
thermal fluctuations. In the following $R^q_{tt'}=\eta^{-1} e^{- q^2
  (t-t')/\eta}$  
(respectively $C^q_{tt'}$)
denote the bare
response function (respectively correlation),
i.e. in the absence of disorder. We now specify the
time regime $t,t'$. 


In the first situation studied below, i.e.
an interface very near the depinning threshold,
we set $T=0$ in (\ref{eq_dyn}). 
Numerical simulations of a manifold of internal size $L$ 
are typically performed on a cylinder, i.e. with
periodicity in $u \sim u+W$ with $W \sim L^\zeta$.
After a time empirically of order fixed number of
turns, i.e. $\tau_L \sim L^z$ according to scaling ($z$ the dynamical
exponent), the system 
reaches a unique global time-periodic steady state
which has been extensively studied \cite{rosso,allemands}. 
Here we are interested instead
in the dynamical regime {\it before} this steady state is established,
i.e. in the limit of times $t$, $t'$ very large compared
to microscopic time scales $a^z$, and such that:
\begin{eqnarray}
t, t' \ll L^z, \xi^z
\end{eqnarray}
where $\xi \sim |f-f_c|^{- \nu}$ is the scale
above which interface motion becomes uncorrelated,
i.e. we mainly focus on the critical regime of 
highly correlated avalanche motion (Fig. 1).
In the computation of the correlations
(\ref{def_C}) and 
response (\ref{def_R}), we will be interested in the scaling limit $q/\Lambda_0
\ll 1$, keeping the scaling variables
\begin{eqnarray}\label{scal_var}
w = q^z(t-t') \quad u = t/t'
\end{eqnarray}
fixed. In the context of the depining transition, the domain of validity
of our approach is depicted in Fig. 1.

To study the second situation, i.e. the relaxational
dynamics of the $d=2$ Cardy-Ostlund model, we set $f=v=0$
in (\ref{eq_dyn}), $\Delta(u)$ being 
a periodic function of period unity. The super-rough
glass phase \cite{tsai_super_rough} for $T<T_g$ is, within RG, 
described by a line of finite temperature fixed points (FP).
Recently, we have obtained analytically \cite{schehr_co_pre}
${\tilde{\cal C}}^{{q}}_{tt'}$ (\ref{def_C}) under the scaling form:
\begin{eqnarray}\label{scaling_cc}
{\tilde{\cal C}}^{{q}}_{tt'} = \frac{T}{q^2} \left(\frac{t}{t'}
\right)^{-\frac{\lambda-d+2}{z}} F_{\tilde{C}}(q^{z_{\text{CO}}} (t-t'),t/t') 
\end{eqnarray}
where $F_{\tilde{C}}(w,u)$ is a universal \cite{foot_non_univ} scaling
function, which was recently confirmed by numerics
\cite{schehr_co_num} in a wider range of temperature. And 
although the numerically 
measured exponents $z_{\text{CO}}$ and  
$\lambda$ were found to be in good agreement with one loop RG
predictions near $T_g$, significant deviations were found to occur at
lower $T$. In addition, the
(connected) structure factor 
${\tilde{\cal C}}^q_{tt}$ is obtained from Eq. (\ref{scaling_cc}) in 
the limit $w \to 0$, $u \to 1$ keeping $w/(u-1) = q^{z_{\text{CO}}}
t$ fixed. Therefore the
dynamical exponent $z_{\text{CO}}$
associated to {\it equilibrium} fluctuations coincides
with the one associated to {\it non-equilibrium} relaxation, which was
actually numerically computed in Ref. \cite{schehr_co_num}.

In both cases Eq. (\ref{eq_dyn}) is studied using the
standard dynamical (disorder averaged) MSR action. 
Correlations (\ref{def_C}) and response
(\ref{def_R}) are obtained as functional derivatives of the 
dynamical {\it effective} action $\Gamma[u,\hat u]$.
It is perturbatively computed
\cite{schehr_co_pre} using the Exact RG equation (ERG) associated to the
multi-local operators expansion introduced in \cite{chauve,scheidl}
and extended to non-equilibrium dynamics in
\cite{schehr_co_pre}. The ERG equations are obtained by varying
an infrared (large scale) cutoff $\Lambda_l= \Lambda_0 e^{-l}$ introduced
in the bare response and correlation functions
($R \to R_{l}$, $C \to C_l$). The information about
non equilibrium dynamics is contained in the
interacting part of~$\Gamma$:
\begin{equation} 
\Gamma_{int}= \int_{xt} i \hat u_{xt} F_{lt}[u_{x}] -
\frac{1}{2} \int_{xtt'} i \hat u_{xt} i \hat u_{xt'}
\Delta_{ltt'}(u_{xt} - u_{xt'}) 
\end{equation} 
where only the solution of the ERG equation to lowest order 
in $\Delta\equiv \Delta_l$ and to one loop is needed here. It reads:
\begin{eqnarray} 
&& \frac{\delta F_{lt}[u_{x}]}{\delta u_{xt'}}  = R^{x=0}_{ltt'} \Delta''(u_{xt} - u_{xt'}) \\
&& \Delta_{ltt'}(u) = \Delta(u) 
 + \Delta''(u)(C^{x=0}_{ltt'} - \frac{1}{2}
  C^{x=0}_{ltt} - \frac{1}{2} C^{x=0}_{lt't'}) \nonumber
\end{eqnarray} 
As in \cite{schehr_co_pre} we need however the FRG equation for
the ''statics'' part to one loop and next order. 
It has the standard form:
\begin{eqnarray}\label{frg_one_loop} 
&& \partial_l \tilde{\Delta}(u) = (\epsilon - 2 \zeta) \tilde{\Delta}(u)
+ \zeta u \tilde{\Delta}'(u)
+ \tilde T \tilde{\Delta}''(u) \nonumber
\\
&& - \frac{1}{2} [ (\tilde{\Delta}(u)  - \tilde{\Delta}(0))^2 ]'' 
\end{eqnarray}
where one defined the rescaled, dimensionless disorder
$\tilde{\Delta}_l(u) = S_d \Lambda_l^{- \epsilon} \Delta_l(u)$
and temperature $\tilde T_l = S_d \Lambda_l^{d-2} T/c$
with $\epsilon=4-d$. The response ${\cal R}^q_{tt'}$ and correlation
function ${\cal 
  C}^q_{tt'}$ at the fixed point ($l \to \infty$) 
are given, to one loop, by the equations:
 \begin{equation}
(\eta \partial_t + q^2  ) {\cal R}^q_{tt'}  = \int_0^t dt_1
   \Sigma_{tt_1} {\cal R}^q_{tt'} 
- \int_0^t dt_1 \Sigma_{tt_1} {\cal R}^q_{t_1t'} \label{eq_R} 
\end{equation}
\begin{equation}
{\cal C}^{ {q}}_{ {t} {t'}} =
2T\int_{t_i}^{{t'}} dt_1 {{\cal R}}^{ {q}}_{ {t}t_1}{{\cal
R}}^{ {q}}_{ {t'}t_1} 
+ \int_{t_i}^{ {t}} dt_1 \int_{t_i}^{ {t'}} dt_2{{\cal
R}}^{ {q}}_{ {t} t_1} D_{t_1t_2}
{{\cal R}}^{ {q}}_{ {t'} t_2} \label{eq_C}
\end{equation}
where the fixed point self-energy is $\Sigma_{tt'}= \frac{\delta
  F_{lt}[u_{x}]}{\delta u_{xt'}}|_{u=0}$ 
and the disorder-noise kernel
$D_{tt'}= \Delta_{ltt'}(u=0)$ are local in space to this order of computation.

We first apply the above equations to the case 
of the depinning transition, just above threshold
$f=f_c^+$. In that case one further introduces
a small finite velocity $v \to 0^+$ in the
above equations (shifting $u_{xt} \to u_{xt} + vt$
evrywhere). Eq (\ref{frg_one_loop}), setting $T=0$ reaches the
standard one loop FP for depinning transition
with $\zeta_{\text{dep}}=\epsilon/3$. This FP function 
being non-analytic at $u=0$,
this results in $\Sigma_{tt'}=R^{x=0}_{ltt'} \Delta''(0^+)$
using the limit $v=0^+$. 
We can now solve the equation for 
${\cal R}^q_{tt'}$ (\ref{eq_R}), perturbatively in the disorder,
as in \cite{schehr_co_pre}, and obtain a solution consistent
with the scaling form
\begin{eqnarray}\label{scal_resp_dep}
{\cal R}^q_{tt'} = \left(\frac{t}{t'}\right)^{\theta_R} q^{z-2} F_R(q^z
(t-t'),t/t')  
\end{eqnarray}
with $z - 2 = - \tilde \Delta^{* \prime \prime}(0^+)$ and 
the novel exponent $\theta_R$ associated to large time
 off equilibrium relaxation: 
\begin{eqnarray}
\theta_R = - \frac{1}{2}  \tilde \Delta^{* \prime \prime}(0^+) = -
\frac{\epsilon}{9} 
\end{eqnarray}
$F_R(w,u) \equiv F_R(w)$ is a universal \cite{foot_non_univ} scaling
function, whose 
expression is given at one loop order by: 
\begin{equation}\label{FR}
F_R(w) = e^{-w} +  \frac{z-2}{2}((w-1) \Ei{(w)}
e^{-w} + e^{-w} - 1) 
\end{equation}
where $\Ei{(w)}$ is the exponential integral function, with the large
$w$ power law behavior $F_R(w) \propto w^{-2}$. This one loop
scaling form $q^{z-2} F_R(q^z(t-t'))$ (\ref{scal_resp_dep}) 
can be written as the Fourier transform, w.r.t time
variable of $1/(q^2 + \Sigma(i \omega))$ 
with $\Sigma(s) = s^{2/z} + s + {\cal O}(s^2) $. Such scaling
forms (\ref{scal_resp_dep})
arise in the context of critical points \cite{janssen_noneq_rg}.
Solving Eq. (\ref{eq_R}) for any finite Fourier mode $q$, 
one obtains the local response function ${\cal R}^{x=0}_{tt'}$
\begin{eqnarray}\label{local_resp}
{\cal R}^{x=0}_{tt'} = 
\frac{A^0_{\cal R} + A^1_{\cal R}
\ln{(t-t')}}{(t-t')^{1+(d-2)/z}}\left(\frac{t}{t'}\right)^{\theta_R}  
\end{eqnarray}
where $A^0_{\cal R}, A^1_{\cal R}$ are non universal,
$\Lambda_0$-dependent, amplitudes (the
logarithmic corrections coming from the large $w$ behavior of
$F_{\cal R}(w,u)$). At this order, (\ref{local_resp}) is compatible with local scale invariance
arguments \cite{henkel_lsi}. Note however that at this order
(\ref{local_resp}) could also be written as \cite{schehr_rim}
${\cal R}^{x=0}_{tt'} = {A_{\cal R}}\left({t}/{t'}
\right)^{\theta_R}{(t-t')^{-(1+a)}}$ 
with $a \neq 1+(d-2)/z$~: clarifying this point requires higher order calculations,
left for future investigations. 

The correlation function is obtained by solving perturbatively the
equation (\ref{eq_C}), at $T=0$ (thus the bare correlation
$C^{x=0}_{tt'}$ and the connected one $\tilde{\cal C}^q_{tt'}$
vanish). One finds in perturbation to one loop and lowest order that
it is consistent with the scaling form:
\begin{eqnarray}\label{scal_correl_dep}
&&{\cal C}^q_{tt'} = q^{-(d+2 \zeta_{\text{dep}})}\left(\frac{t}{t'}
\right)^{\theta_C} F_C(q^z(t-t'),t/t') \\
&&F_C(w,u) = \tilde\Delta^*(0)(1- e^{-w u/(u-1)}) (1- e^{-w /(u-1)})
\nonumber 
\end{eqnarray}
where a priori $\theta_C = {\cal O}(\epsilon)$ which, as
$F_C(w,u)$ is already of order ${\cal O}(\epsilon)$, requires a second
order calculation \cite{foot_fdr_dep}.      
%

We now focus on the relaxational dynamics (\ref{eq_dyn}) of random
periodic system, at low temperature. We first start by deriving the
low $T$ expression of the {\it equilibrium} dynamical exponent
$z_{\text{CO}}$. It is given by the one loop FRG equation
(\ref{frg_one_loop}) together with $\partial_l \ln \eta_l = -
\tilde{\Delta}''(0) = z_{\text{CO}} - 2$. 
We will specialize to
$d=2$ (we remind $S_2=1/(2\pi)$) thus $\epsilon=2$ and therefore
$\tilde T_l = \tilde T = T/(2 \pi c)$ 
is not flowing.
%
%
In that case, there is a line of fixed points $\tilde{\Delta}_T^*(u)$
indexed by 
temperature, analyzed in Ref.~\cite{chauve_creep_long}, for
$T<T_g$. The transition 
at $T=T_g$ can be analyzed in the Fourier representation
$\tilde{\Delta}(u) = \sum_{n \neq 0} \tilde \Delta_n \cos(2 \pi n u)$, 
the linear part of Eq. (\ref{frg_one_loop}) being $
\partial_l \tilde \Delta_n = 2 (1 - \frac{T}{T_g} n^2 ) \tilde
  \Delta_n + {\cal O}(\Delta^2)$, 
where $T_g=c/\pi$ (i.e. $2 - (2 \pi)^2 \tilde T_g =0$). 
The transition corresponds to the lowest harmonic becoming relevant.
At any temperature the fixed point solution of (\ref{frg_one_loop})
can be written in that case: 
\begin{eqnarray}  
&& u = (3/\epsilon) G(\tilde T,\tilde T+\tilde
    \Delta^*(0)-\tilde \Delta^*(u)) \\
&& 4 y_{\pm} = 3  \tilde \Delta^*(0) + \tilde T \pm \sqrt{3(3 
 \tilde \Delta^*(0) + \tilde T)(\tilde \Delta^*(0) + 3 \tilde T)} \nonumber
\end{eqnarray}
where $G(a,b)=\int_a^b y dy/\sqrt{ (y-\tilde T)(y-y_-)(y-y_+) }$, 
and $y_-<0<\tilde T <y_+$. The condition
$\frac{1}{2} = G(\tilde T,y_+)$
yields $\tilde \Delta^*(0)$ as a function of $T$. As $T\to 0$ this
solution converges to the zero temperature solution
$\tilde \Delta_{T=0}^*(u) = \tilde \Delta_{T=0}^*(0) -
  \frac{\epsilon}{6} u(1-u)$, 
and $\tilde \Delta_{T=0}^*(0)=\epsilon/36$. From the FRG equation
at $u=0$ one has the exact relation for all~$T$:
\begin{eqnarray}\label{rel_all_T} 
&& \tilde  T \tilde \Delta^{* \prime \prime}(0) = - \epsilon \tilde
  \Delta^{*}(0) 
\end{eqnarray}
It implies that as $T \to 0$, $\tilde \Delta^{* \prime \prime}(0) \sim
  - \epsilon \tilde 
  \Delta_{T=0}^{*}(0)/\tilde T$ 
which gives, using $\tilde T = T/(2 \pi^2 T_g)$ the one loop estimate:
\begin{eqnarray}\label{expr_z_co} 
&& z_{\text{CO}} -2 \sim \frac{2 \pi^2}{9} \frac{T_g}{T} \simeq 2.19
  \frac{T_g}{T} 
\end{eqnarray}

We now compute the response ${\cal R}^q_{tt'}$ (\ref{def_R}) and the
correlation ${\cal C}^q_{tt'}$
(\ref{def_C}) for the relaxational dynamics defined by
Eq. (\ref{eq_dyn}). By solving to first order in the disorder the
equations 
(\ref{eq_R},\ref{eq_C}) for the present case, one obtains, in the 
limit $q/\Lambda_0 \ll 1$ keeping $q^z(t-t')$, $t/t'$ fixed, the same result as
for the the depinning (\ref{scal_resp_dep}, \ref{scal_correl_dep})
with the substitution of the 
exponents $z$ and $\theta_R$ by $z_{\text{CO}}$ given by
Eq. (\ref{expr_z_co}) 
and $\theta_{\text{CO}}$ given to one loop, as $T\to 0$ by
\begin{eqnarray}\label{expr_theta_co}
\theta_{\text{CO}} \sim  \frac{\pi^2}{9} \frac{T_g}{T} \simeq 1.09
\frac{T_g}{T}   
\end{eqnarray}

For the present case where $T\neq 0$, the connected correlation
function $\tilde{\cal C}^q_{tt'}$ is non
zero. It is given by an equation exactly similar to
Eq. (\ref{eq_C}) with the substitution of $D_{tt'}$ by $D^c_{tt'}$
given to one loop by $D^c_{tt'} = \Delta''(0) C^{x=0}_{ltt'}$.
By solving perturbatively the equation for $\tilde{\cal C}^q_{tt'}$, we 
find a solution consistent with the scaling form given in
Eq. (\ref{scaling_cc}) where $z_{\text{CO}}$ is given by
(\ref{expr_z_co}) and $\lambda = d$ to the order of our calculation,
in good agreement with the numerics \cite{schehr_co_num}
and where $F_{\tilde{C}}(w,u)$ is a universal \cite{foot_non_univ} scaling
function: 
\begin{eqnarray}\label{FC_complete}
&&F_{\tilde{C}}(w,u) = u(F_{\tilde C}^{\text{eq}}(w) - F_{\tilde
  C}^{\text{eq}}(w 
  \frac{u+1}{u-1})) \\ 
&&+ (z-2) u e^{-w \frac{u+1}{u-1}}\left( 
  \Ei{\left(\frac{2w}{u-1}\right)} -  \ln{\left(\frac{2w}{u-1}\right)}
  - \gamma_E \nonumber
\right) 
\end{eqnarray}
where $F_{\tilde{C}}^{\text{eq}}(w) = -\int_{-\infty}^{w} dw' F_R(w')$ and
$\gamma_E$ is the Euler constant. The exponent $\lambda$ in
Eq. (\ref{scaling_cc}) is defined such that $\lim_{u \to \infty}
F_{\tilde{C}}(w,u) = F_{\tilde{C},\infty}(w)$ with, in the limit $T
\to 0$: 
\begin{eqnarray}\label{cc_asymp}
F_{\tilde{C}\infty}(w) = (2+ \frac{4 \pi^2}{9} \frac{T_g}{T}) F_{R}(w)
\end{eqnarray}
Given the scaling form obtained, and
the discussion below Eq. (\ref{scaling_cc}), it is thus consistent to
compare our results for the equilibrium exponent $z_{\text{CO}}$
(\ref{expr_z_co}) to its value obtained in the numerical simulation of
Ref. \cite{schehr_co_num}. This comparison is shown on Fig. 2. We
compare the numerical results both to the low $T$ expansion of
$z_{\text{CO}}$ 
(\ref{expr_z_co}), extrapolated to all temperatures (FRG2), and to the
full expression of $z_{\text{CO}}$
where the value of $-T \Delta''^*(0)$ is
obtained from the 
numerical solution, at finite $T$ -- although only valid at low $T$ --
of Eq. (\ref{rel_all_T}) (FRG1). Both FRG estimates suggest
a rather good agreement, at low $T$, with numerics.

\begin{figure}
{\includegraphics[angle=-90,width=\figwidth]{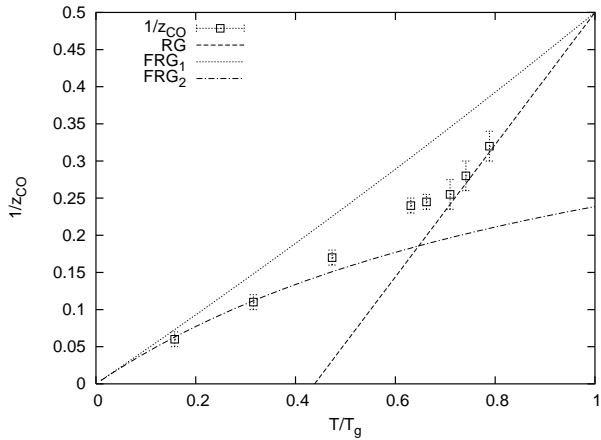}}
\caption{$1/z_{\text{CO}}$ as a function of $T/T_g$. The square symbols are the
 results of the numerical simulation of Ref.\cite{schehr_co_num}. The
 curves FRG1 and FRG2 correspond to our one loop FRG estimate as
 explained in the text. For comparison, we have plotted in dashed line the
 result of the one loop RG calculation in the vincinity of $T_g$
 \cite{tsai_super_rough}.} 
\end{figure}

Finally we can evaluate the non trivial Fluctuation Dissipation Ratio (FDR)
${\cal X}^q_{tt'}$ defined from the {\it connected} correlation
(i.e. such that ${{\cal X}^q_{tt'}} = 1$ at
equilibrium) and found to take the form:
\begin{equation}
\left({{\cal X}^q_{tt'}}\right)^{-1} = {\partial_{t'} \tilde{\cal
    C}^q_{tt'}}/{(T {\cal R}^q_{tt'})} = F_X(q^z(t-t'),t/t') 
 \label{def_FDR}
\end{equation}
In the limit $t/t' \gg 1$, keeping $q^z(t-t')$ fixed, one obtains, in
the low $T$ limit, using (\ref{scal_resp_dep}, \ref{cc_asymp}):
\begin{eqnarray}\label{FX_inf}
\lim_{u \to
  \infty}\frac{1}{{\cal X}^q_{tt'}}
 = 2 + \frac{2 \pi^2}{9} \frac{T_g}{T} = \frac{1}{X_\infty}
\end{eqnarray}  
independently of $w$, a consequence of (\ref{cc_asymp}). Notice also
the identity, given the value of $z_{\text{CO}}$ 
(\ref{expr_z_co}),  
obtained here up to one loop, $z_{\text{CO}}
=\frac{1}{X_\infty}$. This relation was also found in the vicinity of
$T_g$ \cite{schehr_co_pre} and is consitent with numerical simulations
at low $T$ \cite{schehr_co_num}.

In conclusion we have defined and computed new universal exponents
and scaling form for driven interfaces near the depinning transition,
We also showed that the one loop truncation of the FRG yields a good
approximation to numerics for low $T$ aging dynamics of the pinned
periodic manifold in $d=2$. Further numerics near depinning and investigations 
of other predictions of the TBL picture (e.g. barrier
fluctuations), together with more precise RG calculations,
would be of high interest.

GS acknowledges the financial support provided
through the European Community's Human Potential Program 
under contract HPRN-CT-2002-00307, DYGLAGEMEM.

\end{document}